\documentclass[aps,twocolumn,amsmath,amssymb, superscriptaddress]{revtex4}
\usepackage[pdftex]{graphicx}
 \usepackage{amsmath}
\usepackage[T1]{fontenc}
\usepackage{flushend}
\usepackage{amssymb}
\usepackage{amsfonts}
\usepackage{bm}
 \usepackage{amsmath} 
 \usepackage{xcolor}
\usepackage{lipsum}
\usepackage{amsfonts} 
\usepackage{amssymb, mathrsfs}
\usepackage{braket}
\usepackage{graphicx} 
\usepackage{subfigure}
\usepackage{bbm}
\def\beq{\begin{equation}}
\def\eeq{\end{equation}}
\def\bsp{\begin{split}}
\def\esp{\end{split}}
\def\bea{\begin{eqnarray}}
\def\eea{\end{eqnarray}}
\def\ba{\begin{array}}
\def\ea{\end{array}}

\def\dg{\dagger}

\def\lb{\left(}
\def\rb{\right)}

\def\l.{\left.}
\def\r.{\right.}

\def\ra{\rangle}
\def\la{\langle}

\def\bo{\bold{k}}

\begin{document}

\date{\today}
\title{Squeezed Dirac and Topological Magnons in a Bosonic Honeycomb Optical Lattice}
\author{S. A. Owerre}
\affiliation{Perimeter Institute for Theoretical Physics, 31 Caroline St. N.- Waterloo, Ontario N2L 2Y5, Canada}
\author{J.  Nsofini}
\affiliation{Institute for Quantum Computing, University of Waterloo - Waterloo, Ontario N2L 3G1, Canada}
\affiliation{Department of Physics and Astronomy, University of Waterloo - Waterloo, Ontario N2L 3G1, Canada}

\begin{abstract}
Quantum information storage using charge-neutral quasiparticles are expected to play a crucial role in the future of quantum computers. In this regard, magnons or collective spin-wave excitations in  solid-state materials  are promising candidates in the  future of quantum computing.  Here, we study the quantum squeezing of Dirac and topological magnons in a bosonic honeycomb optical lattice with spin-orbit interaction by utilizing the mapping to quantum spin-$1/2$ XYZ Heisenberg model on the honeycomb lattice with discrete Z$_2$ symmetry  and a  Dzyaloshinskii-Moriya  interaction.  We show that the squeezed magnons can be controlled by the Z$_2$ anisotropy and   demonstrate  how the noise in the system is  periodically  modified  in the ferromagnetic and antiferromagnetic phases of the model.  Our results  also apply to solid-state honeycomb (anti)ferromagnetic insulators. 
 \end{abstract}
 \pacs{71.70.Ej,73.23.Ra}
\maketitle

\section{Introduction} 
 Quantum squeezing is the mechanism  for reducing  the noise of a given quantum observable at the expense of enhancing the noise of its conjugate observable \cite{sqq0,sqq,sqq1,sqq2}. The spin squeezing \cite{sqq2a} in particular plays a vital role in the detection of  quantum entanglement \cite{sqq3,sqq4,sqq4a} and  also present itself as a promising candidate for quantum-information processing \cite{sqq5}. In recent years, quantum squeezing  has  expanded tremendously to different systems such as photon \cite{sqq6, sqq7,op,op1} and phonons \cite{sqq8,sqq9}.  
Recently,  squeezed magnons (collective spin-wave excitation)  in solid-state materials have garnered much attention  \cite{jim1,jim,jim2,jim3} as reported in the cubic antiferromagnetic insulators XF$_2$ (X $\equiv$ Mn and Fe) through the impulsive stimulated Raman scattering  \cite{jim1,jim}.  A possible realization in one-dimensional optical lattice which maps to a ferromagnetic spin chain \cite{dong1} has also been proposed \cite{dong}. However, magnon squeezing in ferromagnetic systems necessarily  requires a dipolar interaction \cite{dong,aka, aka1}, which might not be present in some systems. Therefore, it is highly desirable to explore alternative scenarios in which ferromagnetic magnon squeezing can emerge without dipolar interaction and  also the possibility of squeezed magnons in other antiferromagnetic insulators such as the honeycomb antiferromagnetic insulators XPS$_3$ (X $\equiv$ Mn and Fe) \cite{mn1,mn1a,mn2, mn3}.

In recent years, two-dimensional (2D) optical lattices have garnered considerable attention. Atoms trapped in 2D optical lattices  offer a new avenue for understanding the nature of phases in 2D quantum magnetism \cite{dua,por,tro,alt}. In particular, the $p$-orbital bosons trapped in an optical lattice can be used as a model for quantum spin-$1/2$ XYZ Heisenberg model \cite{fer} with a discrete Z$_2$ symmetry. For a particular choice of 2D optical lattice spin-orbit interaction (SOI) or fictitious gauge fluxes can be engineered with laser beams and  provide topologically nontrivial band structures with integer Chern numbers \cite{jot}. In the corresponding quantum spin model, this would correspond to a synthetic Dzyaloshinskii-Moriya (DM) SOI \cite{dm,dm2}, therefore the associated magnetic excitations would correspond to topological magnons. 

Recently, the study of Dirac and topological magnonics in solid-state magnetic systems has come into focus \cite{rc,rc1,lif,sol,frans}. They are expected to open the next frontier of physics, because they are potential candidates towards magnon spintronics and magnon thermal devices \cite{rc2}. On the other hand, magnon qubit and  magnon quantum computing offer a promising avenue for eliminating the difficulties posed by charged electrons \cite{aka1,rc3}. As magnons are charge-neutral quasiparticles, performing quantum computations with  them would possibly require less power than computing with charged electrons \cite{xu} and the information loss through Ohmic heating in electrically  charged quasiparticles would be less in charge-neutral quasiparticles. Therefore magnonic devices would be more efficient in quantum memory and information storage \cite{si,har,spe,kal, haiw}.  The reduction of noise in such a magnonic system indeed requires quantum squeezing of the magnon modes.

In this paper, we study the squeezed  coherent oscillations (periodic modulation of  noise) of Dirac and topological magnons  in a $p$-orbital bosonic atoms trapped in a honeycomb optical lattice. This system maps to a quantum spin-$1/2$ XYZ Heisenberg model  with discrete Z$_2$ symmetry \cite{fer}. We study the magnon squeezing of the corresponding quantum spin system in the ferromagnetic and antiferromagnetic phases  with a DM SOI, which introduces topological features in the associated magnon dispersions. We show  that the squeeze  coherent oscillations of magnons in the ferromagnetic phase requires no dipolar   interactions \cite{dong,aka, aka1}. This is  a consequence of the Z$_2$ symmetry of the Hamiltonian.

 In the antiferromagnetic phase, we  find that the squeeze  coherent oscillations of magnons   depend on the counter-precession of magnon intrinsic spins in the system. Furthermore, we map the system to a Z$_2$-invariant hardcore bosons on the honeycomb lattice and uncover  the mean-field phase diagram with gapped Goldstone modes in each phase. Our results are applicable to  solid-state materials such as  honeycomb antiferromagnetic insulators XPS$_3$ (X $\equiv$ Mn and Fe) \cite{mn1,mn1a,mn2, mn3}. We hope that these results will pave the way towards the utilization of Dirac and topological magnons in quantum information storage and spintronics.

The organization of this paper is as follows. In Sec.~\ref{mod} we introduce the $p$-orbital bosonic atoms trapped in a 2D optical lattice and the mapping to XYZ  quantum spin-$1/2$ Heisenberg model. We also show the symmetry transformations associated with the quantum spin system. In Sec.~\ref{magnon} and Sec. \ref{FM}  we derive the squeeze Hamiltonian of  the XYZ  quantum spin-$1/2$ Heisenberg model on the honeycomb lattice with  DM interaction and discuss the associated magnon band structures.  Sec.~\ref{sque} discusses the squeezing properties  and the  coherent oscillations of  magnon in our model. In Sec.~\ref{conc} we present the concluding remarks.   Appendix \ref{topo} analyzes the topological aspects of magnons in our model and  Appendix \ref{hc} discusses the mapping to Z$_2$-invariant hardcore bosons;  we also uncover   the complete mean-field phase diagram.

\section{ Model }
\label{mod}
The $p$-orbital bosonic atoms of mass $m$ trapped in a 2D optical lattice  can be described by a tight binding Hamiltonian \cite{fer}. At half-filling (zero magnetic field) it  maps to  a  quantum spin-$1/2$ XYZ Heisenberg model 
\begin{align}
\mathcal {\hat H}_{XYZ}&=\sum_{\la i, j\ra}[J\lbrace (1+\gamma)\hat{ S}_{i}^x\hat{ S}_{j}^x+(1-\gamma)\hat{ S}_{i}^y\hat{ S}_{j}^y\rbrace+J_z\hat{ S}_{i}^z\hat{ S}_{j}^z],
\label{model1}
\end{align}
where the symbol $\la i, j\ra$  represents the sum over nearest neighbour (NN) sites and $J,J_z>0$ are exchange constants and $\gamma\neq 0$ is an anisotropy.     The most important feature of the $p$-orbital Bose system is the manifestation of  Z$_2$ symmetry. In the spin language, this corresponds to the transformations $\hat{S}_{ij}^{x}\to-\hat{S}_{ij}^{x}$, $\hat{S}_i^{y}\to-\hat{S}_{ij}^{y}$ and $~\hat{S}_{ij}^{z}\to-\hat{S}_{ij}^{z}$ for $\gamma\neq 0$. 
This is synonymous with the fact that no spin component commutes with the Hamiltonian.   Note that the sign of $\gamma$ in Eq.~\eqref{model1} can be changed by the canonical transformation    $\hat S_{ij}^{x} \to -\hat S_{ij}^{y},~\hat S_{ij}^{y,z} \to \hat S_{ij}^{x,z}$, that is $\pi/2$-rotation about the $z$-axis. Therefore the ground state of  Eq.~(\ref{model1}) is independent of the sign of $\gamma$.  

The mapping from the bosonic $p$-orbital atoms  in a 2D optical lattice to quantum spin $1/2$ XYZ system makes no assumptions regarding the geometry of the 2D lattice \cite{fer}. Here, we study this model on a honeycomb lattice. In the limit $J_z<J(1+\gamma)$ the spins would prefer to anti-align along the $x$-axis and the other terms in Eq.~\eqref{model1} act as quantum fluctuations. In this paper, we work in this limit and set $J_z=J$ and $0< \gamma< 1$, which  preserves the Z$_2$ symmetry of the Hamiltonian. The quantization axis will be chosen along the $x$-direction. However, since a $\pi/2$ rotation about the $y$-axis transforms $\hat S_{ij}^x \to \hat S_{ij}^z$ and $\hat S_{ij}^z \to -\hat S_{ij}^x$, the quantization axis can equally be chosen along the $z$-axis after the transformation.   The XYZ Heisenberg model can also be mapped to Z$_2$-invariant hardcore bosons (see Appendix ~\ref{hc}).

On the honeycomb lattice  spin-orbit interaction (SOI) can be allowed.   In the quantum spin  language with $x$-axis as the quantization axis the linear order term in the perturbative  expansion of the SOI corresponds to the DM interaction    
\begin{align}
\mathcal {\hat H}_{so}=\Delta_{so}\sum_{\la \la i,j\ra\ra} {\nu}_{ij}\hat{x}\cdot\hat{\bf S}_{i}\times\hat{\bf S}_{j},
\end{align}
where $\la\la i, j\ra\ra$ represents sum over next-nearest neighbour (NNN) sites and ${\nu}_{ij}=\pm $ depending on the hopping along the NNN sites. For magnetic insulators, the SOI term is present  due to lack of inversion symmetry  of the lattice  according to  the Moriya rules \cite{dm2}. This occurs on the NNN sites for the honeycomb lattice \cite{sol}.  In the bosonic language, the SOI maps to a fictitious gauge flux   for the bosons, which is anologous to a bosonic version of the Haldane model \cite{fdm}. The presence of gauge flux makes the system topologically nontrivial and can be controlled by  laser beams \cite{jot}.  Note that the SOI  also preserves the Z$_2$ symmetry of the original Hamiltonian \eqref{model1}. Hence, the total quantum spin Hamiltonian can be written as \begin{align}
\mathcal {\hat H}=\mathcal {\hat H}_{XYZ}+\mathcal {\hat H}_{so}.
\label{model2}
\end{align}

\subsection{Antiferromagnetic phase at  half-filling}
\label{magnon}
\begin{figure}
\centering
\includegraphics[width=.8\linewidth]{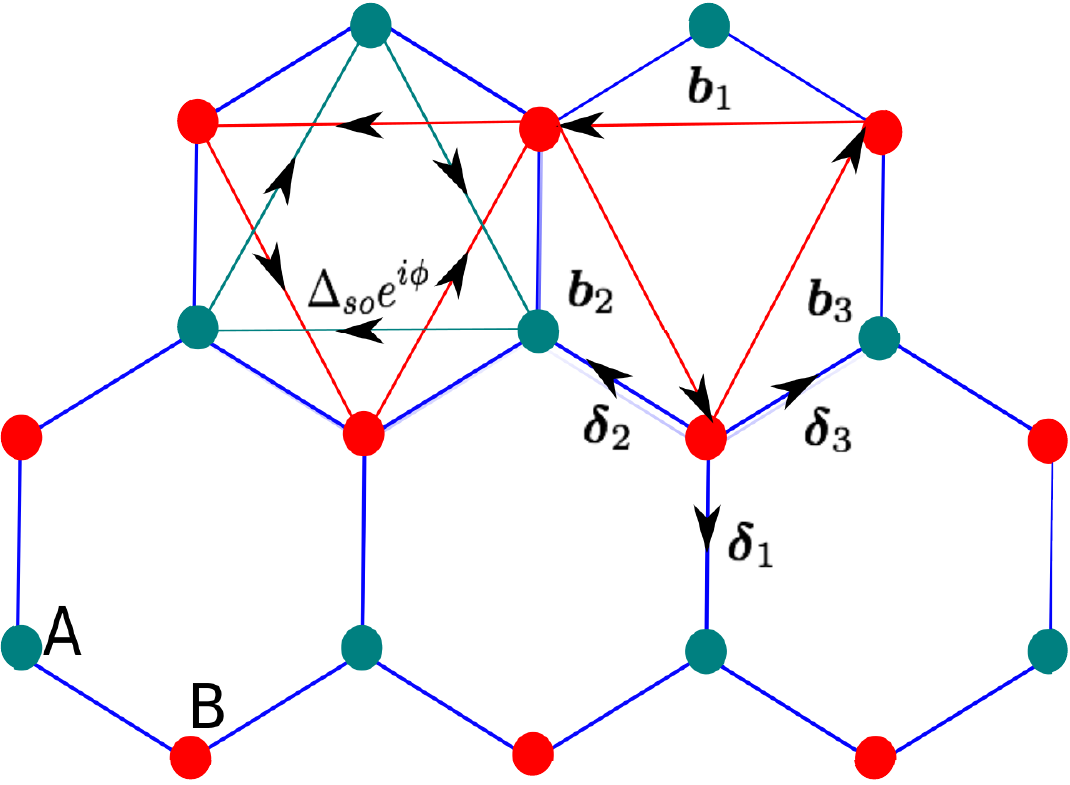}
 \caption{Color online. Schematics of the honeycomb lattice with the gauge flux ($\phi=\pi/2$) treading the NNN bonds. Here, $\boldsymbol{\delta}_i$ and $\boldsymbol{b}_i$ are the vectors connecting the NN and NNN sites respectively. $ \boldsymbol{\delta}_3=a(\sqrt{3}\hat x,~\hat y)/2$, $ \boldsymbol{\delta}_2=a(-\sqrt{3}\hat x,\hat y)/2$ and $ \boldsymbol{\delta}_1=a(0, -\hat y)$. $\bold b_1=-\sqrt{3}a\hat x;~ \bold b_2=a(\sqrt{3}\hat x, -{3}\hat y)/2$. The sublattices $A$ and $B$ are labeled by different colors.}
\label{lat}
\end{figure}

 In this section, we commence with the antiferromagnetic phase at half-filling (zero  magnetic field).  In this model there is no geometric spin frustration and the quantum fluctuations about the mean-field ground state can be represented by the standard Holstein-Primakoff (HP) transformations \cite{hp}:  $
 \hat S_{iA}^{x}= S-  \hat n_{i,A}, ~  \hat S_{iA}^{+}\approx \sqrt{2S} \hat b_{iA}, ~ \hat S_{i A}^{ -}= (S_{iA}^{+})^\dg $;  $\hat S_{iB}^{x}= -S + \hat n_{i,B}, ~  \hat S_{iB}^{+}\approx \sqrt{2S}\hat b_{iB}^\dg, ~\hat S_{iB}^{ -}=(S_{iB}^{+})^\dg$, where $ \hat n_{i\alpha}=\hat b_{i\alpha}^\dagger \hat b_{i\alpha}$;  $\alpha=A,B$ sublattices of the honeycomb lattice in Fig.~\eqref{lat} and $S^\pm=S^z\pm iS^y$ are the raising and lowering spin operators respectively.   We  substitute  the HP transformations  into Eq.~\eqref{model2} and drop the constant mean-field energy. After Fourier transform, the  Hamiltonian in momentum space can be written as 

\begin{align}
&\mathcal{\hat H}= S\sum_{\bo,\alpha,\beta}\bigg[\Omega_{\alpha\beta} \hat b_{\bo \alpha}^\dagger \hat b_{\bo \beta} +\frac{\Delta_{\alpha\beta}}{2} \lb \hat b_{\bo \alpha}^\dagger \hat b_{-\bo \beta}^\dagger +\hat b_{-\bo \alpha} \hat b_{\bo \beta}\rb\bigg],
\label{main1}
\end{align}

where
\begin{align}
\Omega_{\alpha\beta}=
\begin{pmatrix}
v_0-m_\bo& v_1 \lambda_\bo^*\\
v_1 \lambda_\bo &v_0-m_\bo
\end{pmatrix}_{\alpha\beta},
\label{mat1}
\end{align}

\begin{align}
\Delta_{\alpha\beta}=
\begin{pmatrix}
0& v_2 \lambda_\bo^*\\
v_2 \lambda_\bo &0
\end{pmatrix}_{\alpha\beta},
\label{mat2}
\end{align}

\begin{align} 
\lambda_\bo&=\sum_{j=1}^3 e^{i{\bf k}_j\cdot\boldsymbol{\delta}_j},~
m_\bo =2\Delta_{so}\sum_{j=1}^3\sin\bo_j\cdot\bold{b}_j.
\end{align}
The coefficients are given by
\begin{align}
v_1&= \frac{J\gamma}{2},~
v_2= J(1-\gamma/2),~v_0=3JS(1+\gamma).
\end{align}
In order to diagonalize the Hamiltonian \eqref{main1}, a first step would be  to find the eigenvalues of Eqs.~\eqref{mat1} and \eqref{mat2}: $
\Omega_{\alpha\beta}= \Omega_{\bo\alpha}\delta_{\alpha\beta}, ~\Delta_{\alpha\beta}= \Delta_{\bo\alpha}\delta_{\alpha\beta},$
where
\begin{align}
\Omega_{\bo\alpha}&=v_0+ m_\bo +(-1)^\alpha|v_1 \lambda_\bo|,\\
\Delta_{\bo\alpha}&= (-1)^\alpha|v_2 \lambda_\bo|,
\label{de}
\end{align}
with $\alpha=1,2$ for $A,B$ sublattices respectively.
Now, Eq.~\eqref{main1} can be written as
\begin{align}
&\mathcal{\hat H}=\frac{1}{2}\Big(\mathcal{\hat H}_0 + \mathcal{\hat H}_S\Big),
\label{main2}
\end{align}
where
\begin{align}
\mathcal{\hat H}_0&=\sum_{\bo,\alpha}\lb \Omega_{\bo\alpha} \hat b_{\bo \alpha}^\dagger \hat b_{\bo \alpha}+\Omega_{-\bo\alpha} \hat b_{-\bo \alpha}^\dagger \hat b_{-\bo \alpha}\rb,\\
\mathcal{\hat H}_S&=\sum_{\bo,\alpha}\Delta_{\bo\alpha} \lb \hat b_{\bo\alpha}^\dagger \hat b_{-\bo \alpha}^\dagger +\hat b_{-\bo \alpha} \hat b_{\bo \alpha}\rb,
\end{align}
and $\Omega_{\bo\alpha}\neq \Omega_{-\bo\alpha}$ for $\Delta_{so}\neq 0$.
The magnon pair are correlated with wave vectors of equal magnitude but opposite in direction $\bo$ and $-\bo$.  They constitute a squeezed Hamiltonian with two-magnon modes. As shown  in Sec.  \ref{FM} the off-diagonal term is nonzero even in the ferromagnetic phase due to the Z$_2$ symmetry of the Hamiltonian. Now, Eq. \eqref{main2}  can be brought to a diagonal form by the Bogoliubov transformation
\begin{align}
\begin{pmatrix}
\hat  b_{\bo\alpha}\\
\hat  b_{-\bo\alpha}^\dg
\end{pmatrix}=
\mathcal{P}_{\bo\alpha}
\begin{pmatrix}
\hat  d_{\bo\alpha}\\
\hat  d_{-\bo\alpha}^\dg
\end{pmatrix}
\label{bogo}
\end{align}
 where $ \hat d_{\bo\alpha}^\dg( \hat d_{\bo\alpha})$ are the creation (annihilation) operators of the quasiparticles. They obey the commutation relation $[ \hat d_{\bo\alpha},  \hat d_{\bo^\prime\alpha^\prime}^\dg]=\delta_{\bo,\bo^\prime}\delta_{\alpha,\alpha^\prime}$, if $|u_{\bo\alpha}|^2-|v_{\bo\alpha}|^2={\bf I}_{N\times N}$. 
   $\mathcal P_{\bo\alpha}$  is the  paraunitary operator given by
 \begin{align}
 \mathcal P_{\bo\alpha}=\begin{pmatrix}
 u_{\bo\alpha} & -v_{\bo\alpha}\\
-v_{\bo\alpha}^* & u_{\bo\alpha}^*
\end{pmatrix},
\label{bogo1}
 \end{align}
 and satisfy the relation  $\mathcal{P}_{\bo\alpha}^\dg \boldsymbol{\tau}_3 \mathcal{P}_{\bo\alpha}= \boldsymbol{\tau}_3$, where $\boldsymbol{\tau}_3=\text{diag}[{{\bf I}_{N\times N},-{\bf I}_{N\times N}}]$.
The quantities $u_{\bo\alpha}=\text{diag}(u_{\bo 1},u_{\bo 2})$, $v_{\bo\alpha}=\text{diag}(v_{\bo 1},v_{\bo 2})$ can be expressed as
 \begin{align}
u_{\bo\alpha}&=e^{i\phi_\bo}\cosh{\theta_{\bo\alpha}},\quad
v_{\bo\alpha}=\sinh\theta_{\bo\alpha},
\end{align}
and 
\bea
\phi_{\bo}=-\phi_{-\bo}=\tan^{-1}\bigg[\frac{\mathfrak{Im} \lambda_\bo}{\mathfrak{Re} \lambda_\bo}\bigg].
\eea
Here, $\mathfrak{Re}$ and $\mathfrak{Im}$ denote the real and imaginary parts. 
 \begin{align}
\tanh 2\theta_{\bo\alpha}=\frac{\Delta_{\bo\alpha}}{\Omega_{\bo\alpha}}.
\label{the}
 \end{align}
 
 The band structures of magnon are discussed in Appendix \ref{topo}. In the antiferromagnetic phase at half filling, the magnon bands are doubly degenerate at the SU(2) rotationally symmetric point $\gamma=0$  with a Dirac node protected by the linear Goldstone  mode at $\bo =0$ with energy $\omega_{\bo=0}=0$. In the Z$_2$-invariant  phase $\gamma\neq 0$ the  linear Goldstone (Dirac) mode at $\bo =0$ is gapped ($\omega_{\bo=0}\neq 0$) and the degeneracy of the magnon bands is also lifted by  Z$_2$ symmetry of the Hamiltonian.  In this case the Dirac nodes appear at the corners of the Brillouin zone ${\bf K}_\pm=(\pm 4\pi/3/\sqrt{3},0)$ at finite energy regardless of the SOI. As we will show later, the squeezed coherent oscillations of magnons  in the antiferromagnetic phase is independent of the SOI, but depend on the degenerate (at $\gamma=0$) and non-degenerate (at $\gamma \neq 0)$   up and down spins of the propagating magnons on the two sublattices.

 \subsection{Ferromagnetic phase}
\label{FM}
The fully polarized (FP) ferromagnetic phase can be obtained from the antiferromagnetic phase in the presence of large magnetic field applied along the quantization axis. In this case the Hamiltonian can be modeled ferromagnetically at zero field by
\begin{align}
\mathcal {\hat H}_{XYZ}&=-J\sum_{\la i, j\ra}[  (1+\gamma)\hat{ S}_{i}^x\hat{ S}_{j}^x+(1-\gamma)\hat{ S}_{i}^y\hat{ S}_{j}^y +\hat{ S}_{i}^z\hat{ S}_{j}^z],
\label{appen1}
\end{align}
where $J>0$ and $0<\gamma<1$.   We apply the standard Holstein-Primakoff (HP) transformations \cite{hp}: $
\hat S_{i,\alpha}^{ z}= S-   \hat n_{i,\alpha}, ~  \hat S_{i,\alpha}^{  +}\approx \sqrt{2S} \hat b_{i,\alpha}, ~\hat S_{i,\alpha}^{ -}=(\hat S_{i,\alpha}^{  +})^\dg$.
With the inclusion of  $\mathcal {\hat H}_{so}$ the resulting momentum space Hamiltonian is given by

\begin{align}
&\mathcal{\hat H}= S\sum_{\bo,\alpha,\beta}\bigg[\Omega_{\alpha\beta} \hat b_{\bo \alpha}^\dagger \hat b_{\bo \beta} +\frac{\Delta_{\alpha\beta}}{2} \lb \hat b_{\bo \alpha}^\dagger \hat b_{-\bo \beta}^\dagger +\hat b_{-\bo \alpha} \hat b_{\bo \beta}\rb\bigg],
\label{apmain}
\end{align}

where
\begin{align}
\Omega_{\alpha\beta}=
\begin{pmatrix}
v_0-m_\bo& -J_a\lambda_\bo^*\\
-J_a\lambda_\bo &v_0+m_\bo
\end{pmatrix}_{\alpha\beta},
\label{ap1}
\end{align}

\begin{align}
\Delta_{\alpha\beta}=-J_b
\begin{pmatrix}
0& \lambda_\bo^*\\
\lambda_\bo &0
\end{pmatrix}_{\alpha\beta}.
\label{ap2}
\end{align}
We see that the ferromagnetic phase contains off-diagonal magnon modes due to Z$_2$ symmetry of the Hamiltonian.  Diagonalizing  Eqs.~\eqref{ap1} and \eqref{ap2} gives
\begin{align}
\Omega_{\bo\alpha}&=v_0+ (-1)^\alpha\sqrt{m_\bo^2 +|J_a\lambda_\bo|^2},\\
\Delta_{\bo\alpha}&= (-1)^\alpha|J_b\lambda_\bo|,
\end{align}
where  $J_a=J(1-\gamma/2),~J_b=J\gamma/2,$ and $v_0=3J(1+\gamma)$. 
Now, Eq.~\eqref{apmain} can be written as Eq.~\eqref{main2} with
\begin{align}
\mathcal{\hat H}_0&=\sum_{\bo,\alpha}\Omega_{\bo\alpha}\lb \hat b_{\bo \alpha}^\dagger \hat b_{\bo \alpha} +\hat b_{-\bo \alpha}^\dagger \hat b_{-\bo \alpha}\rb,\\
\mathcal{\hat H}_S&=\sum_{\bo,\alpha}\Delta_{\bo\alpha} \lb \hat b_{\bo\alpha}^\dagger \hat b_{-\bo \alpha}^\dagger +\hat b_{-\bo \alpha} \hat b_{\bo \alpha}\rb.
\end{align}

 At $\gamma=0$ the ferromagnetic state is an exact eigenstate of the Hamiltonian  with an average magnetization of $S=1/2$. However,   for $\gamma\neq 0$  the ferromagnetic state is no longer an exact eigenstate of the Hamiltonian due to Z$_2$ quantum fluctuations  emanating from $S^+S^++S^-S^-$ terms,  which reduce the  average magnetization from the  classical value $S=1/2$.  We note that in the usual SU(2) rotationally  invariant ferromagnets ($\gamma= 0$) or U(1)-invariant ferromagnets  the off-diagonal term $\Delta_{\bo\alpha}$ is zero. It can only be induced when the dipolar interaction is taken into account \cite{aka}. In the present model, however, $\Delta_{\bo\alpha}$ is nonzero  provided  $\gamma\neq 0$. The $\Delta_{\bo\alpha}$  term is the hallmark of magnon squeezing in magnetic systems. 
 

We have discussed the topological aspects of magnon in Appendix \ref{topo}. In the ferromagnetic phase the magnon bands  form  Dirac nodes at ${\bf K}_\pm=(\pm 4\pi/3/\sqrt{3},0)$  when SOI is neglected with a quadratic  Goldstone mode at $\bo =0$  for $\gamma=0$ and a gapped  Goldstone mode at $\bo =0$  for $\gamma\neq 0$. Unlike the antiferromagnetic phase, topological magnons are present in the ferromagnetic phase once  SOI is introduced.

\section{squeezing of Magnon}
\label{sque}
\subsection{Magnon squeezed states}
Having derived the two-magnon modes squeezed Hamiltonian, we now turn to the squeezing properties of magnons in our model. The Z$_2$ symmetry of the Hamiltonian ({\it i.e.} $\gamma\neq 0$) provides an interesting squeezing property in this system. We note that up to an irreverent phase factor, the quasiparticle transformation in Eq.~\eqref{bogo} can be written as $\hat d_{\bo\alpha}= \mathcal S_{\bo,-\bo}(z_{\bo\alpha})\hat b_{\bo\alpha}\mathcal S^{\dg}_{\bo,-\bo}(z_{\bo\alpha})$
where  $\mathcal S_{\bo,-\bo}(z_{\bo\alpha})=\exp\big[z_{\bo\alpha}^*\hat b_{\bo \alpha}\hat b_{-\bo \alpha}-z_{\bo\alpha} \hat b_{\bo \alpha}^\dg \hat b_{-\bo \alpha}^\dg \big]$, and $z_{\bo\alpha}= \theta_{\bo\alpha}e^{-i\phi_\bo}$. The unitary operator $\mathcal S_{\bo,-\bo}(z_{\bo\alpha})$ with the property $\mathcal S^{\dg}_{\bo,-\bo}(z_{\bo\alpha})=\mathcal S^{-1}_{\bo,-\bo}(z_{\bo\alpha})$ is called a two-magnon mode squeeze operator similar to  that of phonons \cite{op,op1}. 

The distinguishing feature of the present model is that the Z$_2$ symmetry of the Hamiltonian makes the quasiparticle transformations to be well-defined even for the $\bo=0$ mode. The two-magnon mode vacuum is given by $\ket{0}_{\bo\alpha}\otimes\ket{0}_{-\bo\alpha}$ with  $b_{\bo\alpha}\ket{0}_{\bo\alpha}=b_{-\bo\alpha}\ket{0}_{-\bo\alpha}=0$.  Applying the squeezed operator to a vacuum gives  a squeezed vacuum  defined as $\ket{\psi_\alpha}_0=\mathcal S_{\bo,-\bo}(z_{\bo\alpha})\ket{0}_{\bo\alpha}\otimes\ket{0}_{-\bo\alpha}$, where $d_{\bo\alpha}\ket{\psi_\alpha}_0=d_{-\bo\alpha}\ket{\psi_\alpha}_0=0$. Therefore the  quasiparticle excitations $\hat d_{\pm\bo\alpha}$ are generated by squeezing the $\hat b_{\pm\bo\alpha}$. Hence, they are called two-mode magnon squeezing operators.  We can define a squeezed magnon entangled state  as
\begin{align}
\ket{\Psi}=c_1\ket{\psi_A,\psi_B}_{0}+c_2\ket{\psi_A+1,\psi_B-1}_{0},
\end{align}
where $|c_1|^2+|c_2|^2=1$. 
An entangled magnon state of this form can be utilized in  quantum memory \cite{si,har,spe,kal, haiw}. We also note that the total $x$-component of the spins carried by the magnons $\hat S^x=\sum_i(\hat S_{i,A}^x+\hat S_{i,B}^x)=\sum_i(-\hat{n}_{i,A}+\hat{n}_{i,B})$ is not a conserved quantity for $\gamma\neq 0$. In the HP spin-boson mapping it is easily shown that $\braket{\psi_A|\hat{S}_x|\psi_A}_0=-1$ and $\braket{\psi_B|\hat{S}_x|\psi_B}_0=1$. Therefore the two-magnon  modes in the $A$ and $B$ sublattices carry equal and opposite non-degenerate spins precessing along the $x$-quantization axis for $\gamma\neq   0$. The ideas of  magnon qubit, magnon spintronics, and magnon quantum computing are based on the manipulation of these intrinsic magnon spins.

The squeezing of the magnetization components can be calculated in the squeezed  vacuum states. For instance the variance (squared uncertainty)  of the $y$ and $z$  components of the magnetization can be written as
\begin{align}
\braket{\Delta S_{\alpha;y,z}^{ 2}}_0=\braket{ S_{\alpha;y,z}^{2}}_0-\braket{S_{\alpha;y,z}}_0^2,
\end{align}
 where the average of the magnetization along the $y$ and $z$ directions vanish, $\braket{S_{\alpha;y,z}}_0=0$. For the $\bo=0$ mode we find
 \begin{align}
 \braket{\Delta S_{\alpha;y,z}^2}_0/\mathcal{N}=\frac{1}{4}\exp[{\pm 4\theta_{0\alpha}}],
 \end{align}
where $\theta_{0\alpha}$ is real as given in Eq.~\eqref{the},  $\mathcal{N}$ is the total number of sites and $\pm$ sign applies to $y$ and $z$ components respectively. For ferromagnet (FM) and antiferromagnet (AFM) in mode 1 on sublattice A we obtain
\begin{align}
\theta_{01}^{\text{FM}}&=\frac{1}{2}\tanh^{-1}\lb -\frac{1}{3}\rb,\\
\theta_{01}^{\text{AFM}}&=\frac{1}{2}\tanh^{-1}\lb \frac{-2+\gamma}{2+\gamma}\rb.
\end{align}
In the antiferromagnetic case the squeezing is dependent on $\gamma$, but not in the ferromagnetic case. We see that the  reduction of  the quantum noise  in $z$ component of the  magnetization increases the noise in the $y$ component. 
\begin{figure}
\centering
\includegraphics[width=.9\linewidth]{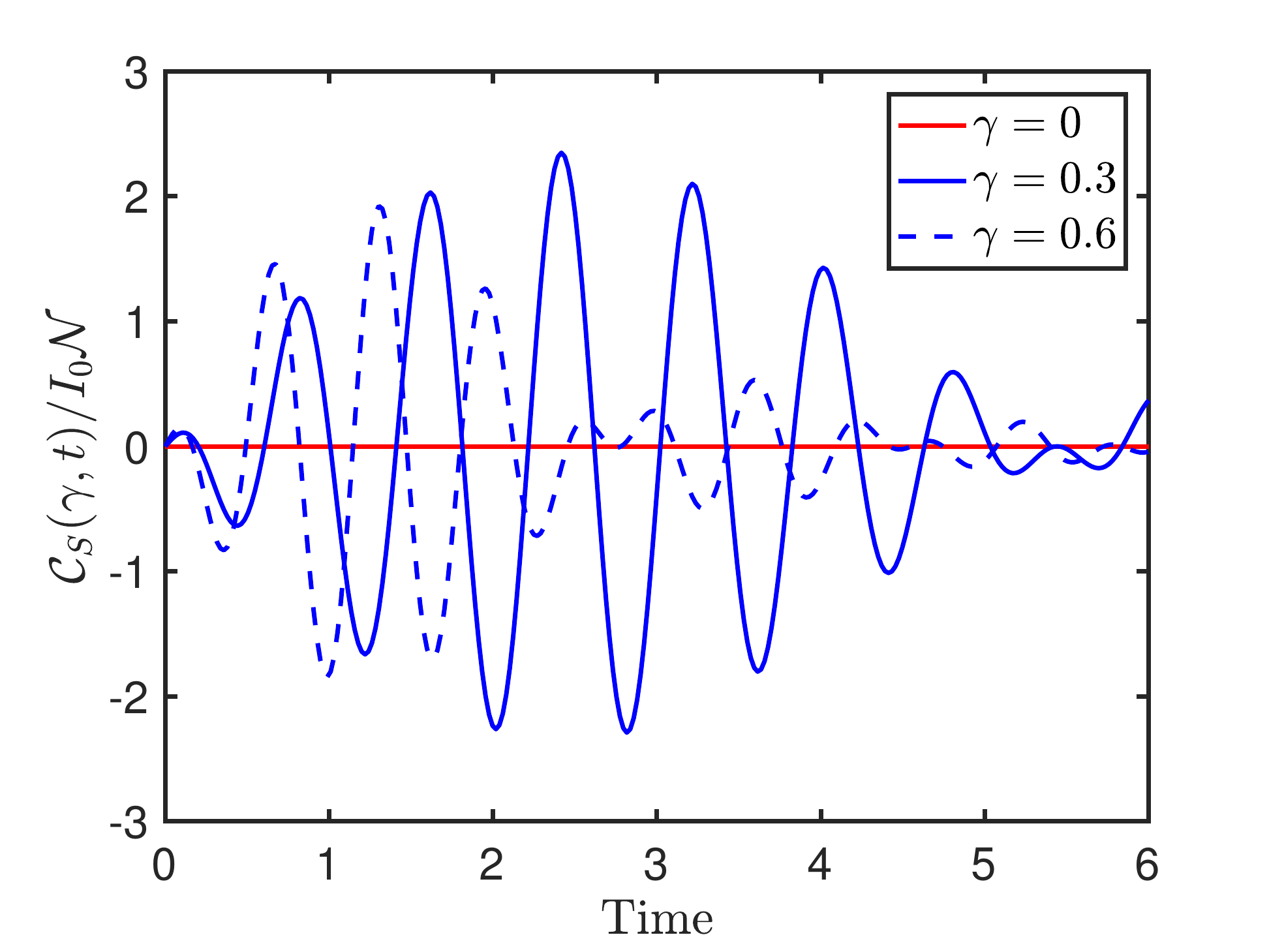}
 \caption{Color online. The symmetric off-diagonal coherent oscillations in the  antiferromagnetic phase as a function of time for several values of $\gamma$. The coherent oscillations are independent of SOI. }
\label{sq1}
\end{figure}

\begin{figure}
\centering
\includegraphics[width=.9\linewidth]{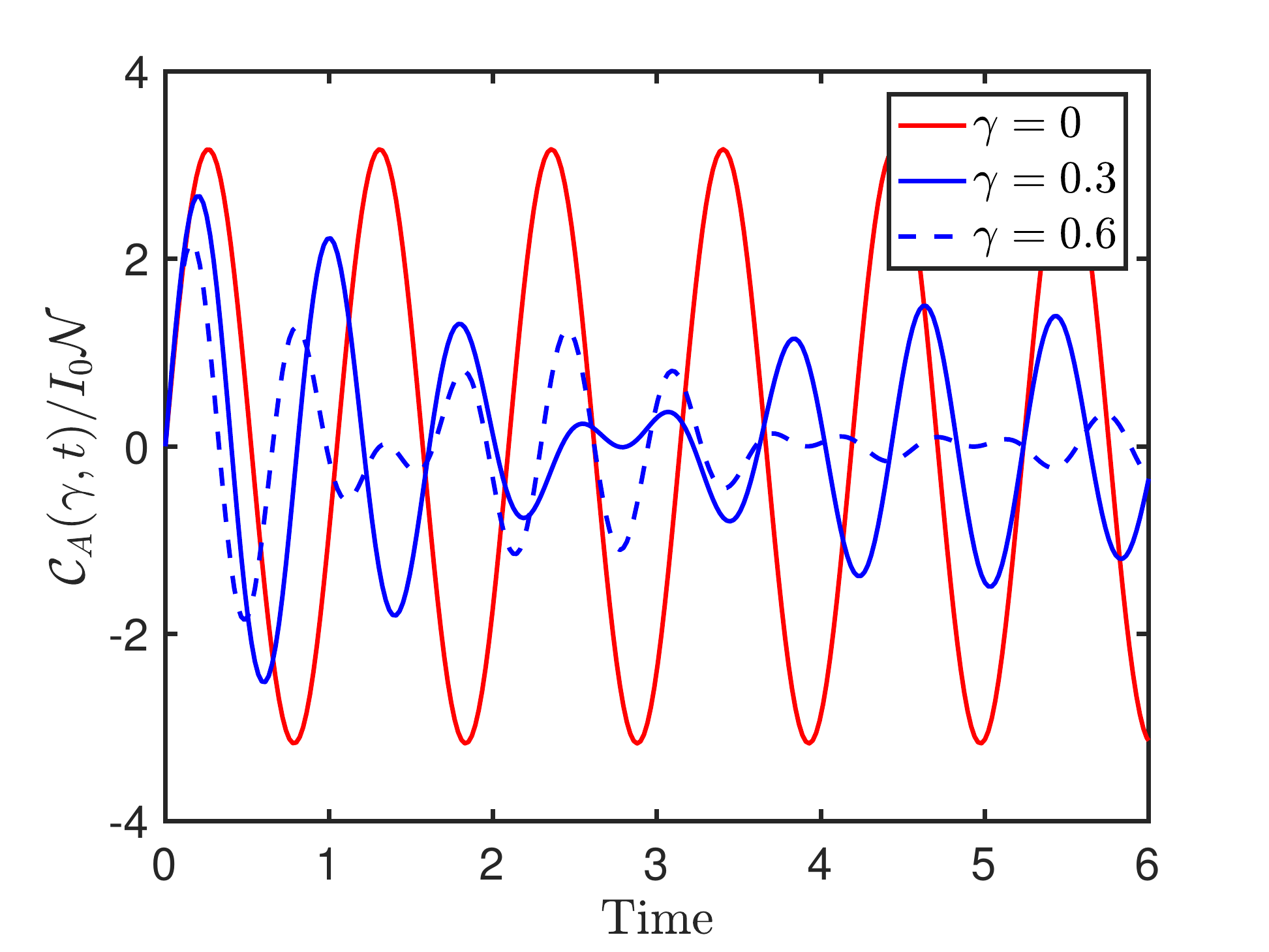}
 \caption{Color online. The antisymmetric off-diagonal coherent oscillations in the  antiferromagnetic phase as a function of time for several values of $\gamma$. The coherent oscillations are independent of SOI. }
\label{sq2}
\end{figure}

\begin{figure}
\centering
\includegraphics[width=.9\linewidth]{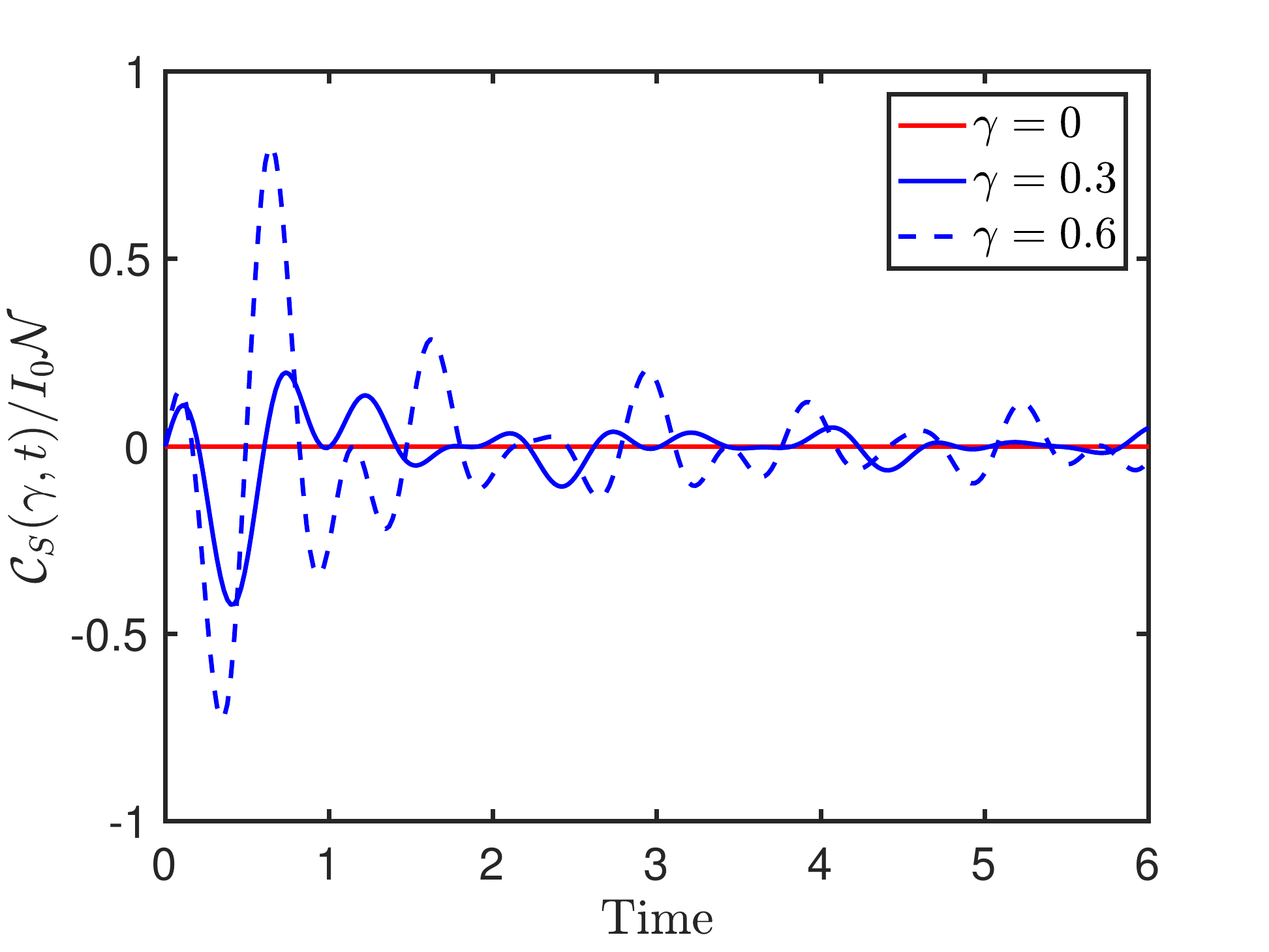}
 \caption{Color online. The symmetric off-diagonal coherent oscillations  in the ferromagnetic phase as a function of time for several values of $\gamma$. The SOI is set to $\Delta_{so}=0.15J$.}
\label{sq}
\end{figure}
\begin{figure}
\centering
\includegraphics[width=.9\linewidth]{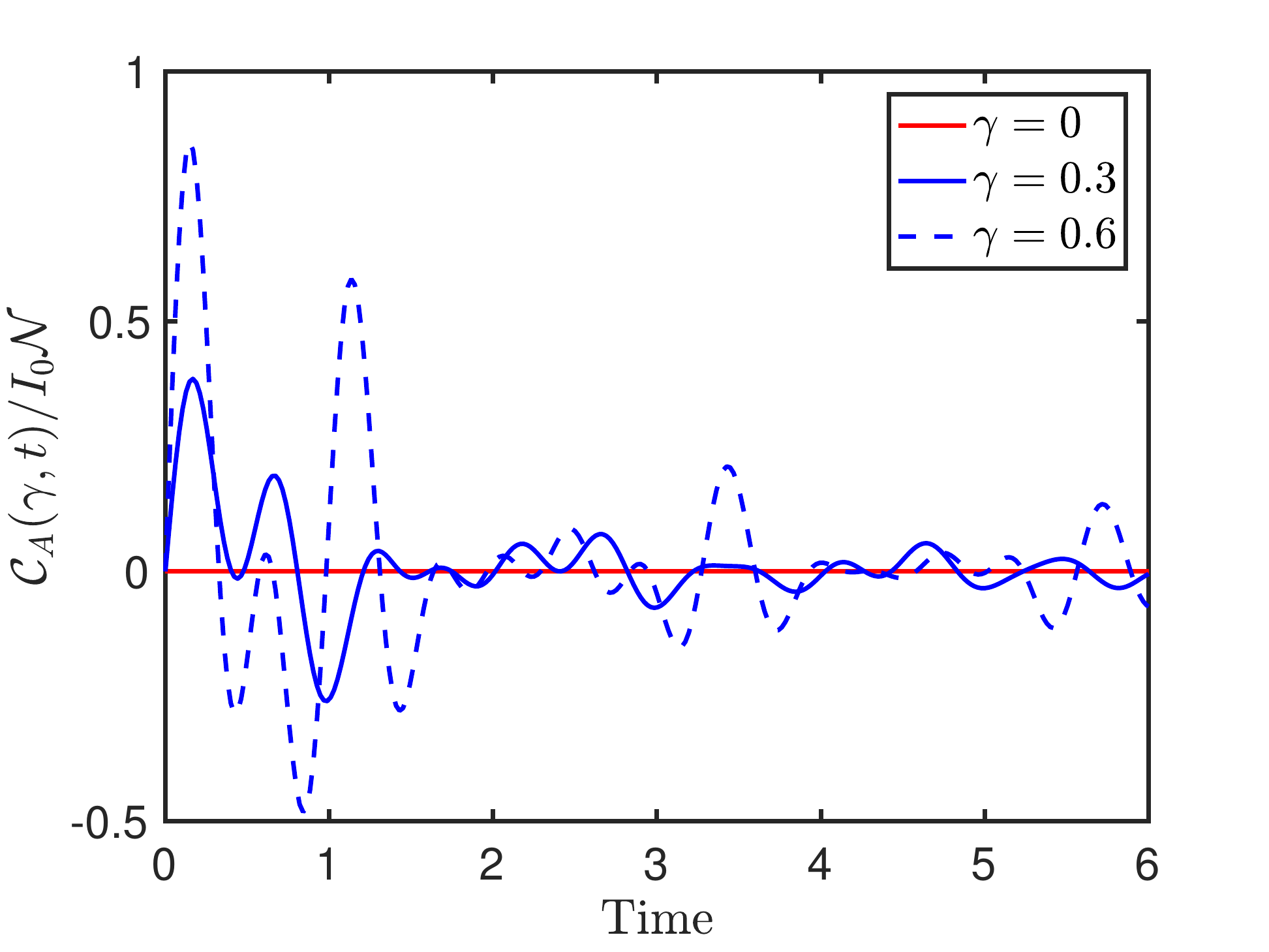}
 \caption{Color online. The antisymmetric off-diagonal coherent oscillations in the  ferromagnetic phase as a function of time for several values of $\gamma$. The SOI is set to $\Delta_{so}=0.15J$. }
\label{sq3}
\end{figure}

\subsection{Coherent oscillations of  squeezed magnons}

In this section, we study the periodic modulation of  noise in the system. We note that the Hamiltonian \eqref{main2} is similar to those  generated through impulsive stimulated Raman scattering between  magnons and light interactions, where a laser pulse is applied on the magnetic insulator \cite{jim,jim1}. We imagine this scenario in a bosonic honeycomb optical lattice or honeycomb  magnetic insulators  \cite{mn1,mn1a,mn2, mn3}.  After the pulse is applied the system will evolve in time  to  new excitations. Suppose a delta function laser pulse  is applied \cite{jim,jim1,gar}, the integration of the Schr\"{o}dinger  equation at $t>0$ gives
\begin{align}
\ket{\psi}_t=e^{i\mathcal {\hat H}_{0}t}\exp\bigg[\sum_{\bo\alpha} \xi_{\bo\alpha}^*\hat b_{\bo \alpha}^\dg \hat b_{-\bo \alpha}^\dg-\xi_{\bo\alpha}\hat b_{\bo \alpha} \hat b_{-\bo \alpha}) \bigg]\ket{\psi_\alpha}_0,
\end{align}
where $\xi_{\bo\alpha}=iI_0\Delta_{\bo\alpha}$, and $I_0$ is a constant that is proportional  to the refractive index and the intensity of the laser pulse. 

In the squeezing of magnons the system should contain  both diagonal and off-diagonal contributions, but  it is the off-diagonal terms, rather than the diagonal ones, which are responsible for the coherent oscillations  in the system. Therefore, we calculate the off-diagonal correlation function of the magnonic operators in the evolved wave function at finite time. We define a symmetric ($S$) and antisymmetric ($A$) off-diagonal correlation functions  generated by 
\begin{align}
\mathcal{C}_S(\gamma,t)&= \sum_{j,\alpha,\beta}\delta_{\alpha\beta}\braket{\hat S_{j,\alpha}^+\hat S_{j,\beta}^+  + \hat S_{j,\alpha}^-\hat S_{j,\beta}^-}_t,\\
\mathcal{C}_A(\gamma,t)&= \sum_{j,\alpha,\beta}\nu_{\alpha\beta}\braket{\hat S_{j,\alpha}^+\hat S_{j,\beta}^+  + \hat S_{j,\alpha}^-\hat S_{j,\beta}^-}_t,
\end{align}
where $\delta_{\alpha\beta}=1$ for $\alpha=\beta$ and  $0$ otherwise;  $\nu_{\alpha\beta}=1$ for $\alpha=\beta\in A$ sublattice and $\nu_{\alpha\beta}=-1$ for $\alpha=\beta \in B$ sublattice. 
The symmetric function  can be regarded as a measure of the Z$_2$ symmetry of the Hamiltonian.
Using the HP transformations and the Baker-Campbell-Hausdorff Lemma we obtain the following expressions
\begin{align}
\mathcal{C}_S(\gamma,t)&=2 I_0 S\sum_{\bo\alpha} \Delta_{\bo\alpha}\sin(2\tilde\Omega_{\bo\alpha}t),\\
\mathcal{C}_A(\gamma,t)&=2I_0S\sum_{\bo\alpha}(-1)^\alpha\Delta_{\bo\alpha}\sin(2\tilde\Omega_{\bo\alpha}t).
\end{align}
In the antiferromagnetic phase, $\tilde\Omega_{\bo\alpha}=(\Omega_{\bo\alpha}+ \Omega_{-\bo\alpha})/2$ and it is independent of the SOI mass term $m_\bo$. The same expression holds for $\mathcal{C}_{S(A)}(\gamma,t)$ in the ferromagnetic phase, but the functions $\Omega_{\bo\alpha}$ and $\Delta_{\bo\alpha}$ are different as given in Sec. \ref{FM} and they also depend on the SOI mass term $m_\bo$.

In Figs.~\ref{sq1} and \ref{sq2} we have shown the symmetric and antisymmetric off-diagonal coherent oscillations in the antiferromagnetic phase respectively. In the former, the coherent oscillations  vanishes at the SU(2) rotationally symmetric point $\gamma=0$, because  the degeneracy of the magnon modes comes with equal and opposite degenerate intrinsic magnon spins and the counter-precession of the spins  cancels each other at the SU(2) symmetric point. However, for $\gamma\neq 0$ the intrinsic magnon spins are no longer degenerate resulting in non-vanishing of the symmetric coherent magnon oscillations.   In the latter,  the two degenerate magnon modes with equal and opposite intrinsic magnon spins at $\gamma=0$ add and the antisymmetric coherent magnon oscillations are nonzero.   As noted above, the SOI does not have any effects on the coherent oscillations in the antiferromagnetic phase.  We note that the SU(2) rotationally symmetric point $\gamma=0$ is a good approximation  to the honeycomb antiferromagnetic insulators XPS$_3$ (X $\equiv$ Mn and Fe) \cite{mn1,mn1a,mn2, mn3}.

The ferromagnetic phase behaves differently from the antiferromagnetic phase as one would expect. In this case the  symmetric and antisymmetric off-diagonal coherent oscillations of magnons are shown in Figs.~\ref{sq} and \ref{sq3} respectively. In contrast to antiferromagnetic phase, they depend on the SOI as well as the Z$_2$ anisotropy $\gamma$.  In this case, the vanishing of the off-diagonal coherent oscillations  at the rotationally symmetric point $\gamma=0$ is not related to the counter-precessions of the magnon intrinsic spins, but due to the fact that  the off-diagonal magnon mode vanishes at $\gamma=0$ (see Sec. \ref{FM}).

\section{Conclusion}
\label{conc}
In this paper, we have shown  that the utilization of Bose atoms in 2D honeycomb optical lattice or equivalently spin-orbit coupling magnetic insulators could play a prominent role in quantum information. We showed that the correspondence between Bose atoms in 2D honeycomb optical lattice and quantum magnetism leads to interesting features. For the $p$-orbital Bose atoms, the discrete Z$_2$ symmetry of the corresponding XYZ quantum spin $1/2$ Hamiltonian on the honeycomb lattice leads to lifted magnon mode degeneracy in the antiferromagnetic phase and gapped Goldstone modes in all phases. In the degenerate modes at the rotationally symmetric point, we found that the coherent oscillations of magnons in the squeezed magnon states depend on the opposite precession of the two-magnon modes with equal and opposite spins. This degeneracy is lifted by a discrete Z$_2$ anisotropy  of the Hamiltonian.  For the  ferromagnetic phase, the Z$_2$ symmetry of the Hamiltonian naturally allows an off-diagonal term necessary for magnon squeezing to exist in stark contrast to rotationally symmetric ferromagnets in which a dipolar  interaction is required \cite{dong,aka, aka1}.  In solid-state materials, spontaneous Raman scattering or femtosecond optical pulses measurements in honeycomb antiferromagnetic insulators XPS$_3$ (X $\equiv$ Mn and Fe) \cite{mn1,mn1a,mn2, mn3} could provide evidence of magnon squeezing similar to those found in the cubic antiferromagnetic insulators XF$_2$ (X $\equiv$ Mn and Fe)  \cite{jim1,jim}.

We also discussed the  Z$_2$-invariant bosonic (magnetic) phases in the hardcore boson mapping (see Appendix \ref{hc}) where the model is devoid of quantum Monte Carlo (QMC) sign problem in  all the  parameter regimes on the honeycomb lattice. We note that the one-dimensional version of the XYZ quantum spin-$1/2$ Hamiltonian  is a playground for exploring quantum entanglement of spin qubit states \cite{can,far}.
\acknowledgements
 Research at Perimeter Institute is supported by the Government of Canada through Industry Canada and by the Province of Ontario through the Ministry of Research
and Innovation.

\appendix

 \section{Topological aspects of magnons}
\label{topo}

\begin{figure}
\centering
\includegraphics[width=1\linewidth]{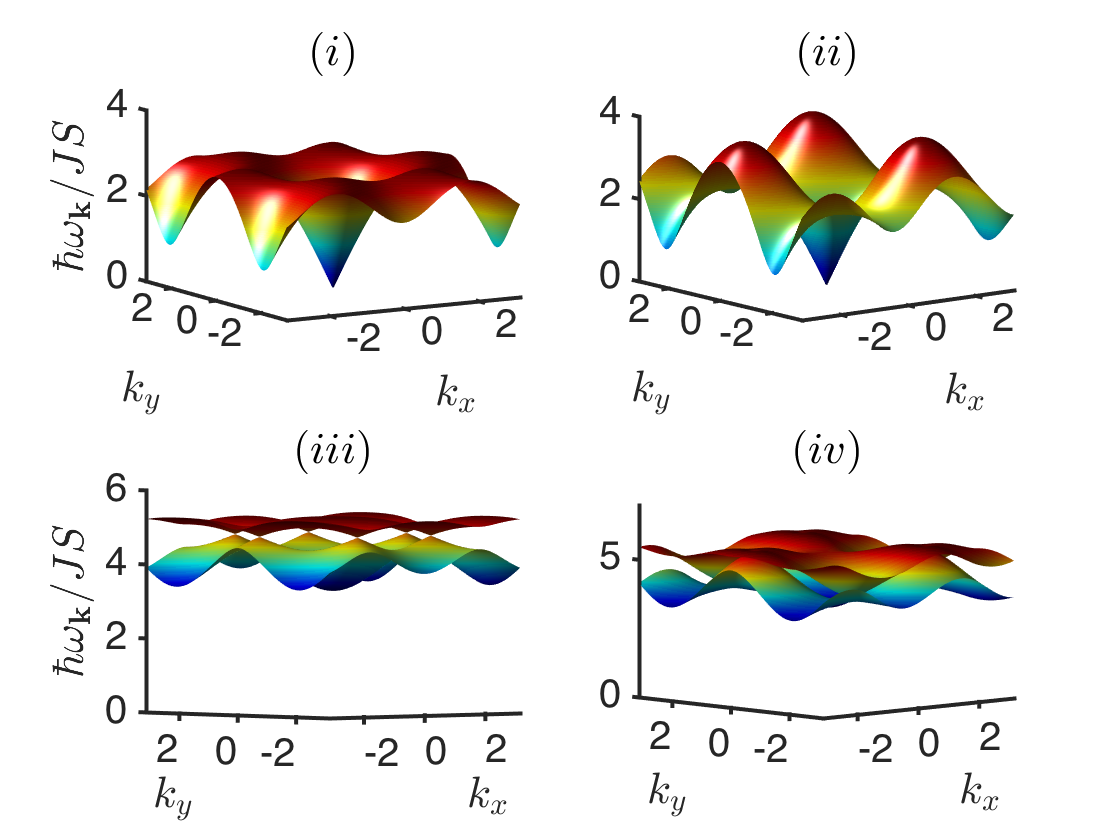}
 \caption{Color online.  Magnon dispersions of the $p$-orbital Bose system on the honeycomb lattice at half-filling.  Top panel. Rotationally symmetric antiferromagnet ($\gamma=0$) without SOI $(\Delta_{so}=0)$ ($i$) and with SOI $(\Delta_{so}=0.15J)$ ($ii$). Bottom panel. Z$_2$ antiferromagnet  ($\gamma=0.6$) without SOI $(\Delta_{so}=0)$ ($iii$) and with SOI $(\Delta_{so}=0.15J)$  $(iv)$.}
\label{band}
\end{figure}

First we study the magnon band structures. At the continuous rotationally symmetric antiferromagnetic point $\gamma=0$ the magnon bands $
 \omega_{\bo\alpha}=\sqrt{\Omega_{\bo\alpha}^2-\Delta_{\bo\alpha}^2}$ are  doubly degenerate and they possess a gapless  linear Goldstone mode near $\bo\to 0$ as expected for a rotationally invariant system as shown  Figs.~\ref{band} $(i)$ and $(ii)$.   For $\gamma\neq0$ the continuous rotational symmetry is broken down to  Z$_2$ symmetry. Quite interestingly,   the degeneracy of the magnon bands is lifted \cite{foot} as well as the Goldstone mode as shown  Figs.~\ref{band} $(iii)$ and ($iv$) with a gap of $\Delta_{\text {gap}}^{\text{AFM}}=3\sqrt{2|\gamma|}$.  This is one of manifestations of the Z$_2$ symmetry of the Hamiltonian \eqref{model1}. However, SOI is unable to open a topological gap at the Dirac points $\bold K_\pm=(\pm 4\pi/3\sqrt{3},0)$, but induces asymmetry in the magnon bands.  The absence of a topological gap is as a result of  the magnetic flux configuration in the  half-filled antiferromagnetic phase. In other words, the DM-induced fictitious magnetic flux is destructive due to  opposite sign of the spins in the N\'eel state.  However, this can be lifted by introducing a moderate external magnetic field perpendicular to the lattice plane.  
 
 \begin{figure}
\centering
\includegraphics[width=1\linewidth]{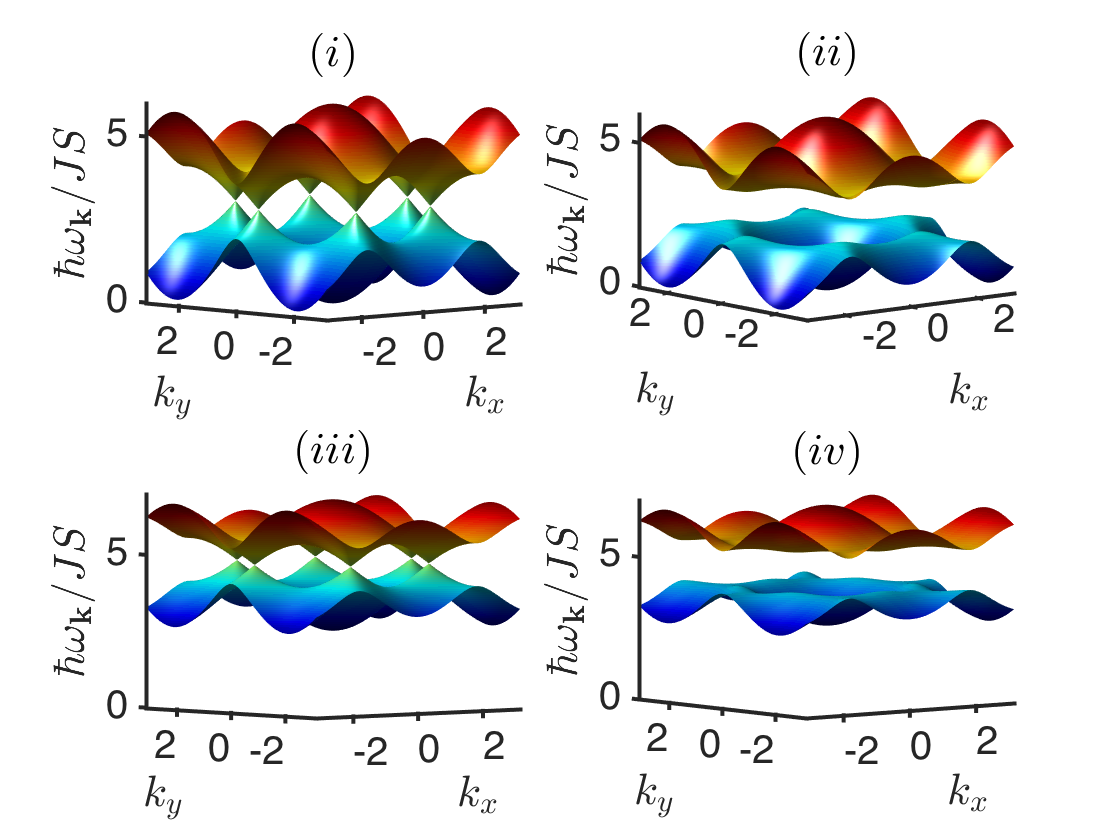}
 \caption{Color online. Magnon dispersions of the $p$-orbital Bose system on the honeycomb lattice at half-filling.  Top panel. Rotationally symmetric ferromagnet ($\gamma=0$) without SOI $(\Delta_{so}=0)$ ($i$) and with SOI $(\Delta_{so}=0.15J)$ ($ii$). Bottom panel. Z$_2$ ferromagnet  ($\gamma=0.2$) without SOI $(\Delta_{so}=0)$ ($iii$) and with SOI $(\Delta_{so}=0.15J)$  $(iv)$.}
\label{band2}
\end{figure}

\begin{figure}
\centering
\includegraphics[width=1\linewidth]{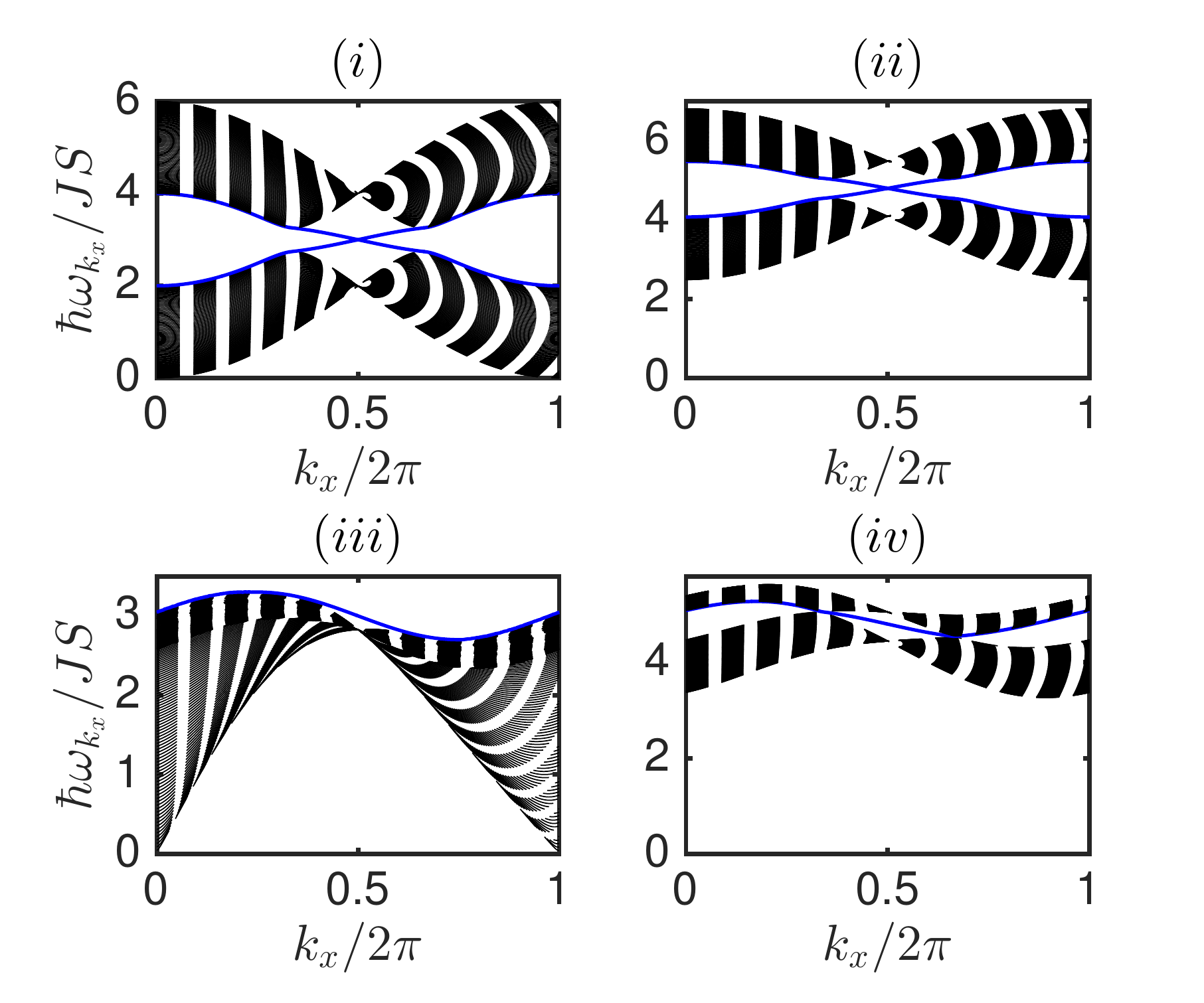}
 \caption{Color online. Top panel: The corresponding zig-zag chiral edge states (blue solid lines)  of Fig.~\eqref{band2} with SOI. Bottom panel: The corresponding zig-zag chiral edge states   of Fig.~\eqref{band} with SOI. The momentum is rescaled in units of $\sqrt{3}$.}
\label{edge}
\end{figure}
 \begin{figure}
\centering
\includegraphics[width=.9\linewidth]{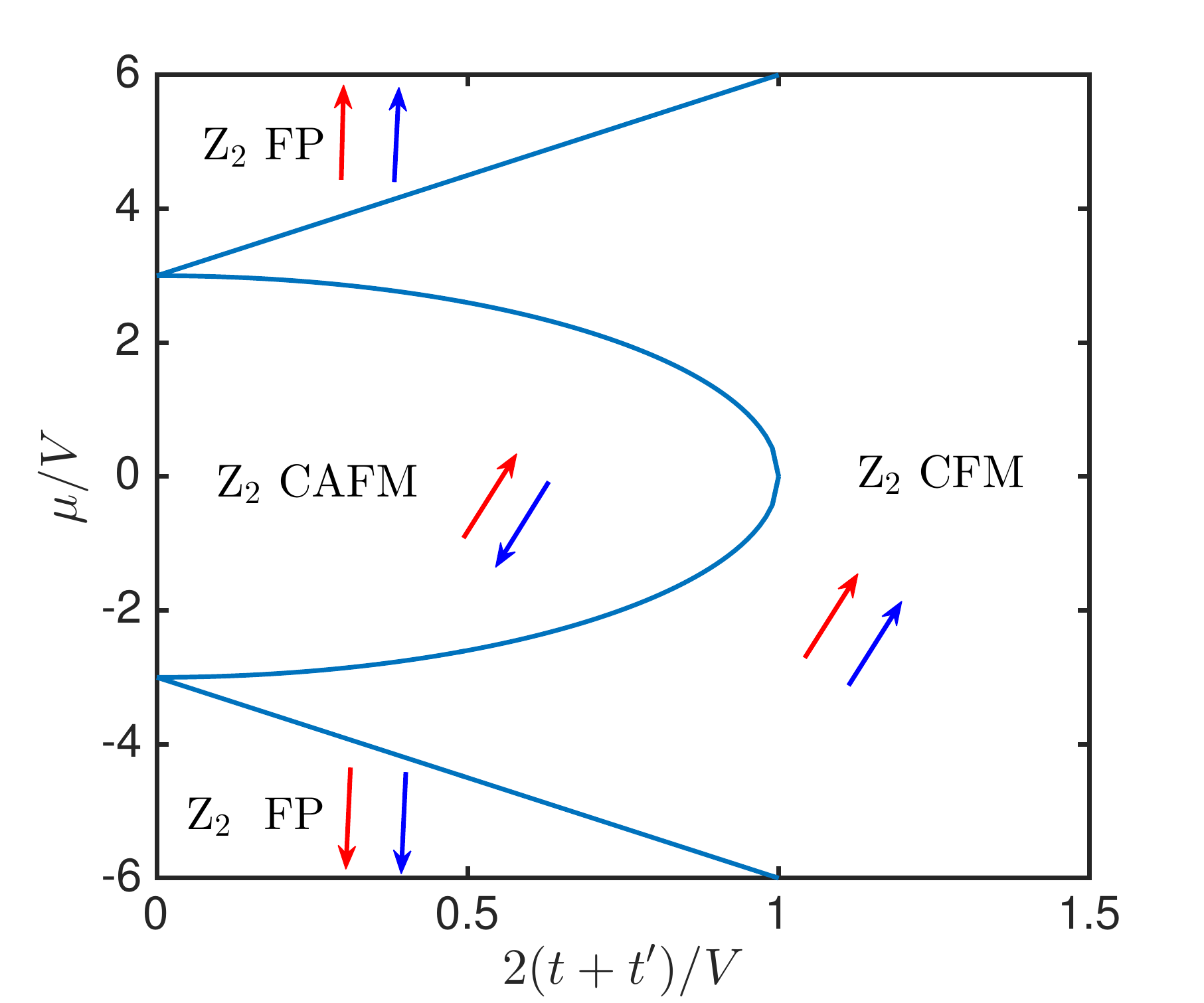}
 \caption{Color online. The mean-field phase diagram of the Z$_2$-invariant  hardcore bosons. Here, FP denotes fully polarized ferromagnet, CAFM is canted antiferromagnet, and CFM is canted in-plane ferromagnet. All the phases have gap Goldstone modes.  The arrows indicate the magnetic spin structure on the two sublattices of the honeycomb lattice in each phase. }
\label{phase}
\end{figure}
 In the ferromagnetic phase which can be achieved from the antiferromagnetic phase by applying a strong magnetic field,  the magnon bosonic operators have off-diagonal terms away from the rotationally symmetric point ($\gamma=0$ ) (see Sec. \ref{FM}).  Therefore, the  quadratic  Goldstone mode near $\bo\to 0$ in Figs.~\ref{band2} $(i)$ and $(ii)$ are gapped ($\Delta_{\text {gap}}^{\text{FM}}=3|\gamma|\sqrt{2}$) by the Z$_2$ symmetry of the Hamiltonian as shown in Figs.~\ref{band2} $(iii)$ and $(iv)$. In this case, however, SOI opens a topological gap at the Dirac points $\bold K_\pm=(\pm 4\pi/3\sqrt{3},0)$.

 The topological aspects of magnons can be studied by defining the Berry curvatures and Chern numbers of the magnon dispersions. The Berry curvature can be defined as
 \begin{align}
\mathcal{B}_{ij,\alpha\bo}=-2\mathfrak{Im}[\boldsymbol{\tau}_3\mathcal (\partial_{k_i}\mathcal P_{\bo\alpha}^\dg)\boldsymbol{\tau}_3(\partial_{k_j}\mathcal P_{\bo\alpha})]_{\alpha\alpha},
 \label{bc1}
 \end{align}
 where $i,j=\lbrace x,y\rbrace$.
We can alternatively write the Berry curvature as
 \begin{align}
\mathcal{B}_{ij;\alpha\bo}=-\sum_{\alpha\neq \alpha^\prime}\frac{2\text{Im}[ \braket{\mathcal{P}_{\bo\alpha}|v_i|\mathcal{P}_{\bo\alpha^\prime}}\braket{\mathcal{P}_{\bo\alpha^\prime}|v_j|\mathcal{P}_{\bo\alpha}}]}{\lb \omega_{\bo\alpha}-\omega_{\bo\alpha^\prime}\rb^2},
\label{chern2}
\end{align}
where   $v_{i}=\partial [\boldsymbol{\tau}_3\mathcal{H}_\bo]/\partial k_{i}$ defines the velocity operators. The Chern number is given by the integration of the Berry curvature over the  momentum space Brillouin zone
\begin{equation}
 n_\alpha= \frac{1}{2\pi}\int_{{BZ}} dk_idk_j~\mathcal{B}_{ij; \alpha\bo}.
\label{chenn}
\end{equation} 
 Topologically, the  top and bottom magnon bands in the ferromagnetic phase carry  Chern numbers of $n_\pm =\pm1$ respectively. Whereas the top and bottom magnon bands in the antiferromagnetic phase at half-filling are topologically trivial with vanishing Chern numbers,  but with  a Berry phase or winding number of $W=\pm 1$ for a closed loop encircling the Dirac nodes for $\gamma\neq 0$. These results are  consistent with the  zig-zag magnon edge modes  in Fig.~\eqref{edge}. The topologically trivial bands have only one chiral edge state connecting the Dirac magnon points in the bulk bands, whereas the topologically nontrivial bands have gapless magnon edge states at the time-reversal-invariant momentum $k_x=\pm\pi/\sqrt{3}$ and $ k_y=0$. 

\section{Hardcore bosons}
\label{hc}
 The quantum spin-$1/2$ XYZ Heisenberg model can be mappped to hardcore bosons. In the limit $J_z=J<J(1+\gamma)$, the transformation has the form $\hat a_i^\dg \leftrightarrow \hat S^{+}_i$, $\hat a_i \leftrightarrow \hat S^{-}_i$, and $\hat n_i \leftrightarrow \hat S^{x}_i+1/2$, where $\hat S_i^\pm=\hat S_{i}^z \pm i\hat S_{i}^y$ and $\hat n_i=\hat a_i^\dg\hat a_i$. They obey the algebra $[\hat a_i, \hat a_j^\dg]=0$ for $i\neq j$ and $\lbrace \hat a_i, \hat a_i^\dg \rbrace=1$. Hence, the  spin-$1/2$ XYZ Hamiltonian maps to the bosonic Hamiltonian 
\begin{align}
 \mathcal{\hat H}_{XYZ}&=t\sum_{\la ij\ra}\lb \hat a_{i}^\dagger \hat a_{j} + h.c.\rb+t^\prime\sum_{\la ij\ra}\lb \hat a_{i}^\dagger\hat a_{j}^\dagger + h.c.\rb \nonumber\\& + V\sum_{\la ij\ra} \hat n_i \hat n_j-\mu\sum_i \hat n_i,
\label{hard}
\end{align}
where the constant terms have been dropped. Here, $t=J(1-\gamma/2)/2,~t^\prime=J\gamma/4,~V=J(1+\gamma),~\mu=H_x$. In the opposite limit $J_z>J(1+\gamma)$, we have that $t=J/2,~t^\prime=J\gamma/2$, and $V=J_z$ with $\mu=H_z$. Therefore, the $p$-orbital bosonic atoms in a 2D optical lattice \cite{fer} can be also be studied in terms of hardcore bosons. We note that unlike frustrated systems the model \eqref{hard} is devoid of the debilitating quantum Monte Carlo (QMC) sign problem in  all the  parameter regimes on the honeycomb lattice. The mean-field phase diagram is depicted in Fig.~\ref{phase}. In the Z$_2$-invariant model, quantum fluctuations are suppressed \cite{owee} due to gapped Goldstone modes. Thus,  we expect that  the mean-field phase diagram  will capture the essential features of the quantum phase diagram. The only difference is that the classical phase boundaries will be  slightly modified.


\begin{thebibliography}{99}
\bibitem{foot}
 In the anisotropic U(1)-invariant collinear antiferromagnets the magnon bands are also doubly degenerate, hence the lifted degeneracy of the magnon bands for $\gamma\neq 0$ is a consequence of the Z$_2$ symmetry and not because of the anisotropy.
\end{thebibliography}
\end{document}